\documentclass[aps,prd,twocolumn,nofootinbib,showpacs]{revtex4-1}

\usepackage{graphicx}

\usepackage[colorlinks=true,linkcolor=blue,citecolor=blue, urlcolor=blue]{hyperref}

\begin{document}

\title{Reconsideration of the QCD corrections to the $\eta_c$ decays into light hadrons using the principle of maximum conformality}

\author{Bo-Lun Du}
\author{Xing-Gang Wu}
\email{wuxg@cqu.edu.cn}
\author{Jun Zeng}
\author{Shi Bu}
\author{Jian-Ming Shen}

\affiliation{Department of Physics, Chongqing University, Chongqing 401331, P.R. China}

\begin{abstract}

In the paper, we analyze the $\eta_c$ decays into light hadrons at the next-to-leading order QCD corrections by applying the principle of maximum conformality (PMC). The relativistic correction at the ${\cal{O}}(\alpha_s v^2)$-order level has been included in the discussion, which gives about $10\%$ contribution to the ratio $R$. The PMC, which satisfies the renormalization group invariance, is designed to obtain a scale-fixed and scheme-independent prediction at any fixed order. To avoid the confusion of treating $n_f$-terms, we transform the usual $\overline{\rm MS}$ pQCD series into the one under the minimal momentum space subtraction scheme. To compare with the prediction under conventional scale setting, $R_{\rm{Conv,mMOM}-r}= \left(4.12^{+0.30}_{-0.28}\right)\times10^3$, after applying the PMC, we obtain $R_{\rm PMC,mMOM-r}=\left(6.09^{+0.62}_{-0.55}\right) \times10^3$, where the errors are squared averages of the ones caused by $m_c$ and $\Lambda_{\rm mMOM}$. The PMC prediction agrees with the recent PDG value within errors, i.e. $R^{\rm exp}=\left(6.3\pm0.5\right)\times10^3$. Thus we think the mismatching of the prediction under conventional scale-setting with the data is due to improper choice of scale, which however can be solved by using the PMC.

\end{abstract}

\maketitle

The heavy quark mass provides a natural hard scale for the heavy quarkonium decays into light hadrons or photons. Calculations of their decay rates are considered as one of the earliest applications of  pQCD. The charmonium has become a popular field since the discovery of $J/\psi$ resonance at SLAC and Brookhaven in 1974. There are lots of successful experimental studies about charmonium, including the precise measurements of spectrum, lifetimes and branch ratios, cf. a comprehensive review given in the PDG~\cite{Olive:2016xmw}. At the same time, many theoretical efforts have been tried for an appropriate description of charmonium. As an important breakthrough, a systematic pQCD analysis of the heavy quarkonium inclusive annihilation and production has been given within the nonrelativistic QCD theory (NRQCD) in 1995~\cite{Bodwin:1994jh}.

According to the NRQCD framework, the quarkonium decay rate can be factored into a sum of products of the short-distance coefficients and the long-distance matrix elements (LDMEs). The short-distance coefficients are perturbatively calculable in a power series of $\alpha_s$. The LDMEs can be estimated by means of the velocity power counting rule, i.e. the LDMEs can be classified in terms of the relative velocity between the constituent quarks of the heavy quarkonium. Especially, the color-singlet ones can be directly related to the wavefunction (derivative of the wavefunction) at the origin, which then can be calculated via proper potential models.

The decay rates of the pseudoscalar quarknium into light hadrons and photons have been calculated at the next-to-leading order (NLO) level~\cite{Barbieri:1979be, Hagiwara:1980nv}. The relativistic corrections at the ${\cal{O}}(\alpha_sv^2)$-order have been given in Refs.\cite{Guo:2011tz, Jia:2011ah}. Within the NRQCD factorization framework, the decay rate of the $\eta_c$ into light hadrons or photons can be expressed as
\begin{eqnarray}
\Gamma(\eta_c \to {\rm LH})  &=&  { F_1({}^1S_0) \over m_c^2} \langle \eta_c| {\cal O}_1({}^1S_0) |\eta_c\rangle \nonumber\\
&& +{G_1({}^1S_0)\over m_c^4}\langle \eta_c| {\cal P}_1({}^1S_0) | \eta_c\rangle  + \cdots
\end{eqnarray}
and
\begin{eqnarray}
\Gamma(\eta_c \to \gamma\gamma) & =&  { F_{\gamma\gamma}({}^1S_0) \over m_c^2} \langle \eta_c| {\cal O}_1({}^1S_0) |\eta_c\rangle \nonumber\\
&& +{G_{\gamma\gamma}({}^1S_0)\over m_c^4}
\langle \eta_c| {\cal P}_1({}^1S_0) |\eta_c\rangle  + \cdots,
\end{eqnarray}
where $F_1$, $G_1$, $F_{\gamma\gamma}$ and $G_{\gamma\gamma}$ are short distance coefficients. The symbol $\cdots$ stands for the contributions from high-dimensional LDMEs which are at least at the level of ${\cal O}(v^4 \Gamma)$. $m_c$ is the $c$-quark pole mass~\footnote{The choice of pole mass avoids the ambiguity of using $\overline{\rm MS}$-mass for separating the renormalization group involved $\beta$-terms~\cite{Wang:2013akk}.}. $v^2$ is the squared heavy quark or antiquark velocity in the meson rest frame. For the case of $\eta_c$, it can be calculated by
\begin{eqnarray}
\langle v^2 \rangle_{\eta_c} ={\langle \eta_c| {\cal P}_1({}^1S_0) |\eta_c\rangle \over  m_c^2 \langle \eta_c| {\cal O}_1({}^1S_0) |\eta_c\rangle}.
\end{eqnarray}
To suppress the uncertainty from the LDMEs, one usually calculates the ratio
\begin{eqnarray}
R &=& \frac{\Gamma(\eta_c \to \mathrm {LH} )}{\Gamma(\eta_c \to \gamma\gamma )} \nonumber \\
   &=& R_0(\mu)\left\{1+\left[2\beta_0\left(\frac{8}{3}+\mathrm{ln}\frac{u^2}{4 m_c^2}\right) -\frac{13\pi^2}{2}+74 \right.\right. \nonumber\\
   &  & \quad\quad \left.\left.+\left(\frac{7\pi^2}{3}-2\beta_0-14\right) \left\langle {v^2} \right \rangle_{\eta_c}\right]a(\mu) + \cdots \right\}, \label{ratio msbar}
\end{eqnarray}
where $a(\mu)={\alpha_s(\mu)}/{(4\pi)}$,  $R_0(\mu)=\frac{81 \pi ^2 C_F}{2 \alpha ^2 N_C} a^2(\mu)$,  $\mu$ is an arbitrary renormalization scale, and $\beta_0=11-\frac{2}{3}n_f$ ($n_f$ being the active flavor number) is the leading $\beta$-term of the renormalization group function. It is noted that the factorization scale dependence is missing at this level, which is the case even at the NNLO level~\cite{Feng:2017hlu}, we are thus free of the factorization scale-setting problem.

It is conventional to take the renormalization scale as the typical momentum flow of the process or the one to eliminate the large logs of the pQCD series, we call this conventional scale-setting approach. As will be shown later, such a simple treatment on scale shall introduce large scale uncertainty and make the low-order prediction unreliable. At present the $\eta$ decays into light hadrons or photons have been calculated up to NNLO level, which still shows a poor pQCD convergence~\cite{Czarnecki:2001zc, Feng:2015uha, Feng:2017hlu}. Thus by simply pursuing higher-and-higher order terms may not be the solution for those high-energy processes. In fact, even if we obtain a small scale uncertainty for global quantities such as the total cross-section or the decay rate at a certain fixed order, it is due to cancelations among different orders; the scale uncertainty for each order is still uncertain and could be very large. Two such examples for Higgs boson decay and the hadronic production of Higgs boson can be found in Refs.\cite{Zeng:2015gha, Wang:2016wgw}. When one applies conventional scale-setting, the renormalization scheme- and initial renormalization scale- dependence is introduced at any fixed order. Thus a proper scale-setting approach is important for fixed-order predictions.

Such large scale uncertainty has been long observed, and to improve the accuracy of $R$, Ref.\cite{Bodwin:2001pt} suggested to resum the final-state chains of the vacuum-polarization bubbles and got $R^{\rm{NNA}}=(3.01\pm0.30\pm0.34)\times10^3$ for naive non-Abelianization resummation~\cite{Beneke:1994qe} and $R^{\rm{BFG}}=(3.26\pm0.31\pm0.47)\times10^3$ for background-field-gauge resummation~\cite{DeWitt:1967ub}, respectively. Both predictions are consistent with the world average given by Particle Data Group (PDG) in year 2000~\cite{Groom:2000in}, which gives $R^{\rm exp}=\left(3.3\pm1.3\right)\times10^3$. As an attempt, the authors of Ref.\cite{Bodwin:2001pt} also presented a prediction by using the Brodsky-Lepage-Mackenzie (BLM) scale-setting approach~\cite{Brodsky:1982gc} and got a much larger $R$ value, i.e. $R^{\rm{BLM}}=9.9\times10^3$.

We should point out that those predictions are different from the value derived from the new experimental measurements, which shows $R^{\rm exp}=\left(6.3\pm0.5\right)\times10^3$~\cite{Olive:2016xmw}. The BLM prediction given in Ref.\cite{Bodwin:2001pt} is questionable. Thus it is interesting to show whether an improved pQCD analysis could be done and could explain the new $R^{\rm exp}$, as is the purpose of this paper. Especially, it is important to show whether the mismatching of the data and the pQCD prediction is caused by improper choice of scale or by some other reasons.

A novel scale-setting approach, the Principle of Maximum Conformality (PMC)~\cite{Brodsky:2011ta, Brodsky:2011ig, Mojaza:2012mf, Brodsky:2013vpa}, has been developed in recent years. The PMC satisfies renormalization group invariance~\cite{Brodsky:2012ms} and it reduces in $N_C\to 0$ Abelian limit~\cite{Brodsky:1997jk} to the standard Gell-Mann-Low method~\cite{GellMann:1954fq}. A more convergent pQCD series without factorial renormalon divergence can be obtained. The PMC scales are physical in the sense that they reflect the virtuality of the gluon propagators at a given order, as well as setting the effective number ($n_f$) of active flavors. The resulting resummed pQCD expression thus determines the relevant ``physical" scales for any physical observable, thereby providing a solution to the renormalization scale-setting problem. Because all the scheme-dependent $\{\beta_i\}$-terms in pQCD series have been resummed into the running couplings with the help of renormalization group equation, the PMC predictions are renormalization scheme independent at every order. Such scheme independence can be demonstrated by using commensurate scale relations~\cite{Brodsky:1994eh} among different observables. A number of PMC applications have been summarized in the review~\cite{Wu:2013ei, Wu:2014iba, Wu:2015rga}. The PMC provides the underlying principal for the BLM, and in the following, we shall adopt the PMC to set the renormalization scale.

Up to NLO level, the expression of $R$ can be rewritten as
\begin{eqnarray}
R = r_{1,0}a^2(\mu) + \left[ r_{2,0} + 2 \beta_0 r_{2,1} \right]a^3(\mu)+{\cal O}(a^{4}),
\end{eqnarray}
where the $\rm{\overline{MS}}$-coefficients $r_{i,j}$ can be read from Eq.(\ref{ratio msbar}), in which $r_{i,0}$ are conformal ones. Following the standard PMC procedures, we get
\begin{eqnarray}
R =r_{1,0}a^2(Q_1)+ r_{2,0}a^3(Q_1),    \label{RPMCMSbar}
\end{eqnarray}
where $\ln {Q^2_1}/{\mu^2} = -{r_{2,1}}/{r_{1,0}}$. We have set the PMC scale $Q_2=Q_1$, whose exact value can be determined by the NNLO term which is not available at present. If directly using the $\overline{\rm MS}$-scheme expression (\ref{ratio msbar}), we shall obtain a small PMC scale $Q_1=0.86$ GeV or $0.78$ GeV for the prediction with or without relativistic correction. It is already close to the low-energy region, this explains why a large $R^{\rm BLM}$ is obtained in Ref.\cite{Bodwin:2001pt}. [At the NLO level, the BLM prediction is the same as the PMC prediction if all $n_f$-terms are pertained to $\alpha_s$-running.] For this case, a reliable prediction can only be obtained by using certain low-energy $\alpha_s$-model, which however will introduce extra model dependence for the prediction.

Following the idea of PMC, only those $\{\beta_i\}$-terms that are pertained to the renormalization of running coupling should be absorbed into the running coupling. For the processes involving three-gluon or four-gluon vertex, the scale-setting problem is more involved~\cite{Binger:2006sj}. Thus to avoid such ambiguity of applying the PMC on $R$, similar to the case of QCD BFKL Pomeron~\cite{Brodsky:1998kn, Zheng:2013uja, Hentschinski:2012kr}, we shall first transform the results from the $\overline{\rm MS}$-scheme to the momentum space subtraction scheme (MOM-scheme)~\cite{Celmaster:1979dm, Celmaster:1979xr} and then apply the PMC.

For the purpose, we adopt the perturbative relation between the $\overline{\mathrm{MS}}$-scheme running coupling and the mMOM-scheme one as~\cite{vonSmekal:2009ae}
\begin{eqnarray}
a^{\rm{\overline {MS} }}(\mu) = a^{\rm{mMOM}}(\mu) \left[ 1  - 4D_{1} a^{\rm{mMOM}} + \cdots \right],  \label{alpha relation}
\end{eqnarray}
where for the Landau gauge, $ D_1 =d_{1,0}+d_{1,1} n_f$, $d_{1,0} =\frac{169}{144}N_C$, and $d_{1,1} =  - \frac{5}{18}$. We then obtain
\begin{widetext}
\begin{eqnarray}
R^{\rm{mMOM}}(\mu) = R_0^{\rm{mMOM}}(\mu)\left\{ {{a^{\rm{mMOM}}}\left[ {2{\beta _0}\left( {\ln \frac{\mu ^2}{4m_c^2} + 1} \right) - \frac{{13{\pi ^2}}}{2} + \frac{{165}}{2}} + \left( {\frac{{7{\pi ^2}}}{3} - 2{\beta _0} - 14} \right)\left\langle {{v^2}} \right\rangle_{\eta_c}  \right] + 1} \right\},
\end{eqnarray}
\end{widetext}
where $R_0^{\rm{mMOM}}(\mu)=\frac{81 \pi ^2 C_F}{2 \alpha ^2 N_C} \left(a^{\rm{mMOM}}(\mu)\right)^2$. After applying the PMC, we obtain a new PMC scale $Q'_{1}=\exp(-3d_{1,1}) Q_{1}$, which equals to $1.99$ GeV or $1.80$ GeV for the prediction with or without relativistic correction. Such a larger PMC scale indicates a reliable pQCD prediction can be achieved by using the mMOM scheme.

To do the numerical calculation, we adopt the $c$-quark and $b$-quark running masses as the $\rm{\overline{MS}}$-scheme ones~\cite{Olive:2016xmw}: $\overline{m}_c(\overline{m}_c)=(1.27\pm 0.03)$ GeV and $\overline{m}_b(\overline{m}_b)=(4.18_{-0.03}^{+0.04})$ GeV. By using the relation between the pole mass $m_Q$ and the $\rm{\overline{MS}}$-scheme running mass $\bar m_Q$~\cite{Gray:1990yh, Broadhurst:1991fy, Chetyrkin:1999ys, Melnikov:2000qh}:
\begin{eqnarray}
m_Q = \bar m_Q\left( \bar m_Q \right)\left[1 + \frac{4\bar \alpha _s\left(\bar m_Q\right)}{3\pi} + \cdots \right],
\end{eqnarray}
we obtain $m_c=1.49\pm0.03$ GeV. To be consistent, we adopt the two-loop $\alpha_s$-running, whose behavior is fixed by using the reference point $\alpha_s(m_Z)=0.1181\pm 0.0011$~\cite{Olive:2016xmw}. And we adopt $\left\langle {{v^2}} \right\rangle_{\eta_c}=0.430{\rm GeV}^2/m_c^2$~\cite{Bodwin:2007fz, Chung:2010vz}.

\begin{table}[htb]
\centering
\caption{Asymptotic scales under the $\rm{\overline{MS}}$-scheme and the mMOM-scheme with different flavor number $n_f$.} \label{lambda}
\begin{tabular}[b]{cccc}
\hline
 & ~~$n_f=3$~~ & ~~$n_f=4$~~ &~~$n_f=5$~~ \\ \hline
 ~~$\Lambda_{\rm{\overline{MS}}}^{(n_f)}$~~ & $0.370^{+0.018}_{-0.018}$ & $0.325^{+0.018}_{-0.018}$ & $0.228^{+0.014}_{-0.014}$ \\
~~$\Lambda_{\rm {mMOM}}^{(n_f)}$~~ & $0.566^{+0.027}_{-0.026}$ & $0.531^{+0.030}_{-0.028}$ & $0.397^{+0.025}_{-0.024}$ \\
\hline
\end{tabular}
\end{table}

Numerical results of the QCD asymptotic scales $\Lambda_{\rm{\overline{MS}}}$ and $\Lambda_{\rm mMOM}$ under Landau gauge are listed in Table~\ref{lambda}, where the errors are dominantly caused by the uncertainty $\Delta\alpha_s(m_Z)=\pm 0.0011$. The asymptotic scales for different schemes satisfy the relation~\cite{vonSmekal:2009ae, Zeng:2015gha}, ${\Lambda_{\rm{mMOM}}}/ {\Lambda_{\rm{\overline{MS}}}} = \exp({2D_1}/{\beta _0})$.

As a cross-check, by using the same input parameters, we obtain the same  $\overline{\rm MS}$-scheme prediction on $R$ under conventional scale-setting as that of Ref.\cite{Bodwin:2001pt}. Due to the reasons listed above, we shall adopt the mMOM-scheme to do our following discussions.

\begin{figure}[htb]
\includegraphics[width=0.48\textwidth]{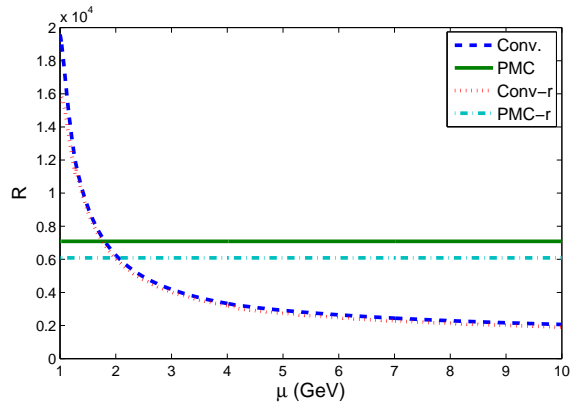}
\caption{The ratio $R$ at the NLO level versus the initial choice of $\mu$ under the mMOM scheme. $m_c=1.49$ GeV. The symbol ``$-r$" stands for relativistic corrections. For conventional scale setting, the sensitivity of $\mu$ is very large. After applying the PMC, $R$ is independent to the choice of $\mu$. }
\label{mu-dependence}
\end{figure}

We present the PMC prediction on $R$ at the NLO level versus the initial choice of $\mu$ in Fig.\ref{mu-dependence}, which is under the mMOM scheme and both the results before and after applying the PMC are presented. Under conventional scale setting, $R$ shows a strong scale dependence which decreases with the increment of $\mu$. More explicitly, by varying $\mu$ from $m_c$ to $4m_c$, the ratio $R$ will change from $\sim9\times10^3$ to $\sim3\times10^3$. After applying the PMC, the PMC scale $Q'_1$ is the same for any choice of $\mu$, leading to scale independent prediction. The relativistic correction brings an extra $\sim 2\%$ contribution to the conventional prediction and $\sim 14\%$ contribution to the PMC prediction. Thus the relativistic correction is important, especially for the PMC predictions. Fig.\ref{mu-dependence} shows that if choosing $\mu=Q'_1$, the values of $R$ under conventional scale setting shall be equal to the PMC ones.

\begin{table}[htb]
\centering
\begin{tabular}[b]{cccccccc}
\hline
 & ~~~LO~~~ & ~~~NLO~~~ &~~~Total~~~ & ~~~$\kappa$~~~  \\
 \hline
~~Conv.~~ & $2.18\times10^3$ & $2.03\times10^3$  & $4.21\times10^3$ & $0.93$ \\
~~Conv.-r~~ & $2.18\times10^3$ & $1.94\times10^3$  & $4.12\times10^3$ & $0.89$ \\
~~PMC~~ & $4.28\times10^3$ &$2.81\times10^3$ & $7.09\times10^3$  & $0.66$ \\
~~PMC-r~~ & $3.66\times10^3$ & $2.43\times10^3$ & $6.09\times10^3$  & $0.67$ \\
\hline
\end{tabular}
\caption{The LO and NLO terms of $R$ under the mMOM-scheme, respectively. $\mu=2m_c$. The symbol ``$-r$" stands for relativistic corrections. $\kappa=R_{\rm{NLO}} / R_{\rm{LO}}$, which shows the relative importance of the NLO-term and the LO-term of $R$. }
\label{lonlo}
\end{table}

After applying the PMC, due to the elimination of divergent renormalon terms as $n! \beta_0^n\alpha_s^n$, the pQCD series shall be more convergent. We present the LO and NLO terms of $R$ before and after applying the PMC in Table \ref{lonlo}. We define a parameter $\kappa=R_{\rm{NLO}} / R_{\rm{LO}}$ to show the relative importance of the NLO-term and the LO-term. Table \ref{lonlo} confirms that a better pQCD convergence can be achieved by applying the PMC. A larger $\kappa$ and a larger scale uncertainty for each term indicate that one cannot get the exact value for each term by using a guessed scale suggested by conventional scale-setting. Analyzing the pQCD series in detail, we observe that the scale errors for conventional scale-setting are rather large for each term, and a possible net small scale error for a pQCD approximant is due to correlations/cancelations among different orders.  On the other hand, due to the fact that the running of $\alpha_s$ at each order has its own $\{\beta_i\}$-series governed by the renormalization group equation, the $\beta$-pattern for the pQCD series at each order is a superposition of all the $\{\beta_i\}$-terms which govern the evolution of the lower-order $\alpha_s$ contributions at this particular order. Thus, inversely, the PMC scale at each order is determined by the known $\beta$-pattern, and the individual terms of $R$ at each order are well determined.

\begin{figure}[htb]
\includegraphics[width=0.48\textwidth]{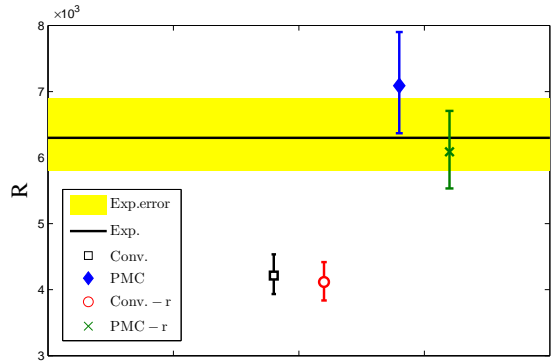}
\caption{Uncertainties of $R$ under the mMOM-scheme from the $c$-quark pole mass $m_c$ and the asymptotic scale $\Lambda_{\rm mMOM}$, where the error bars are squared averages of the errors from those two error sources. The symbol ``$-r$`` stands for corresponding relativistic corrections. The experimental prediction of Ref.\cite{Olive:2016xmw} is presented as a comparison.}
\label{errorbar}
\end{figure}

We present the theoretical uncertainties for the conventional and the PMC scale settings in Fig.\ref{errorbar}, in which the errors are squared averages of the ones from the choices of the $c$-quark pole mass $m_c$ and the asymptotic scale $\Lambda_{\rm mMOM}$. As a comparison, the experimental prediction of Ref.\cite{Olive:2016xmw} is also presented. Under conventional scale-setting, Fig.\ref{errorbar} shows that the errors caused by $m_c$ and $\Lambda_{\rm mMOM}$ is smaller than the case of PMC scale-setting, which is however diluted by the quite large scale uncertainty. For example, the value of $R$ with [or without] relativistic corrections shall be varied within the large region of $\left(4.12^{+4.69}_{-1.50}\right) \times10^3$ [or $\left(4.21^{+5.06}_{-1.56}\right) \times10^3$] for $\mu\in [m_c,4m_c]$. Under PMC scale-setting, the scale uncertainty is greatly suppressed, and the $R$ uncertainty is dominated by the choices of two parameters $m_c$ and $\Lambda_{\rm mMOM}$, which give about $10\%$ contribution to $R$. The value of $R$ decreases with the increment of $m_c$, and increases with the increment of $\Lambda_{\rm mMOM}$. More explicitly, we have
\begin{eqnarray}
 R_{\rm{Conv,mMOM}}=\left(4.21^{+0.13+0.29}_{-0.11-0.26}\right)\times10^3, \\
R_{\rm{Conv,mMOM}-r}=\left(4.12^{+0.12+0.28}_{-0.11-0.26}\right)\times10^3, \\
 R_{\rm{PMC,mMOM}}=\left(7.09^{+0.32+0.75}_{-0.29-0.66}\right) \times10^3,\\
 R_{\rm{PMC,mMOM}-r}=\left(6.09^{+0.21+0.58}_{-0.19-0.52}\right) \times10^3,
 \end{eqnarray}
where the first error is for $m_c\in[1.46, 1.52]$ GeV and the second one is caused by taking $\Lambda_{\rm mMOM}$ to be the values listed in Table~\ref{lambda}.

Fig.\ref{errorbar} shows that the conventional prediction of $R$ with or without relativistic correction is about $3.6\sigma$ derivation from the data. This discrepancy becomes even larger by including the NNLO term~\cite{Feng:2017hlu}, thus the authors there even doubt the validity of NRQCD theory for this particular observable~\footnote{The NNLO results given in Ref.\cite{Feng:2017hlu} cannot be conveniently adopted for setting the PMC scales. We need to confirm which $n_f$-terms are conformal and which are not, thus the scale-setting procedures are much more involved. And, we think such a complex NNLO calculation need to be confirmed by other groups.}. However, Fig.\ref{errorbar} shows that after applying the PMC, the pQCD prediction and the data are consistent with each other within reasonable errors even at the NLO level. This indicates that the large discrepancy between the data and the pQCD prediction is caused by improper choice of renormalization scale, and a simple guessed scale may lead to false prediction or false conclusion. Thus a proper setting of the renormalization scale is important for lower-order predictions.

As a summary, in this paper, we have studied the ratio of the $\eta_c(1S)$ decay rate into hadrons over its decay rate into photons by applying the PMC. The PMC provides a systematic way to set the optimal renormalization scale for high energy process, whose prediction is free of initial renormalization scale dependence at any fixed order. A more convergent pQCD series can be achieved and the residual scale dependence due to unknown high-order terms are highly suppressed. Fig.\ref{errorbar} shows that the large discrepancy between the data and the pQCD prediction by using a guessed scale suggested by conventional scale-setting can be cured by applying the PMC. The PMC, with its solid physical and theoretical background, greatly improves the precision of standard model tests, and it can be applied to a wide variety of perturbatively calculable processes.

\vspace{0.2cm}

\noindent Acknowledgment: This work was supported in part by the National Natural Science Foundation of China under Grant No.11625520. \\

\end{document}